\documentclass[iop]{emulateapj}
\usepackage{amsmath}
\usepackage{graphicx}
\usepackage{rotating}
\usepackage{subfloat}
\usepackage{color}
\usepackage{natbib}
\usepackage{soul}
\setcitestyle{authoryear,round,semicolon,aysep={,},yysep={,},notesep={; }}
\begin{document}
\title{EVIDENCE FOR UBIQUITOUS HIGH-EW NEBULAR EMISSION in $\lowercase{z}\sim7$ GALAXIES: TOWARDS A CLEAN MEASUREMENT of the SPECIFIC STAR FORMATION RATE USING A SAMPLE OF BRIGHT, MAGNIFIED  GALAXIES}
\author{R. Smit\altaffilmark{1}, R. J. Bouwens\altaffilmark{1}, I. Labb{\'e}\altaffilmark{1}, W. Zheng\altaffilmark{2}, 
L. Bradley\altaffilmark{3},
M. Donahue\altaffilmark{4},
D. Lemze\altaffilmark{2},
J. Moustakas\altaffilmark{5},
 K. Umetsu\altaffilmark{6},
  A. Zitrin\altaffilmark{7},
 D. Coe\altaffilmark{3},
M. Postman\altaffilmark{3},
V. Gonzalez\altaffilmark{8},
M. Bartelmann\altaffilmark{7},
N. Ben{\'{\i}}tez\altaffilmark{9},
T. Broadhurst\altaffilmark{10},
 H. Ford\altaffilmark{2},
C. Grillo\altaffilmark{7},
L. Infante\altaffilmark{11}, 
Y. Jimenez-Teja\altaffilmark{9},
S. Jouvel\altaffilmark{12},
D.D. Kelson\altaffilmark{13},
O. Lahav\altaffilmark{14},
D. Maoz\altaffilmark{15},
 E. Medezinski\altaffilmark{2},
P. Melchior\altaffilmark{16},
 M. Meneghetti\altaffilmark{17},
J. Merten\altaffilmark{18},
A. Molino\altaffilmark{9}, 
 L. Moustakas\altaffilmark{18},
 M. Nonino\altaffilmark{19},
P. Rosati\altaffilmark{20}, 
S. Seitz\altaffilmark{22}
}
\altaffiltext{1}{Leiden Observatory, Leiden University, NL-2300 RA Leiden, Netherlands}
\altaffiltext{2}{The Johns Hopkins University}
\altaffiltext{3}{Space Telescope Science Institute}
\altaffiltext{4}{Michigan State University}
\altaffiltext{5}{Siena College}
\altaffiltext{6}{Academia Sinica, Institute of Astronomy \& Astrophysics}
\altaffiltext{7}{Universitat Heidelberg}
\altaffiltext{8}{University of California, Riverside}
\altaffiltext{9}{Instituto de Astrof\'isica de Andaluc\'ia}
\altaffiltext{10}{University of the Basque Country}
\altaffiltext{11}{Universidad Catolica de Chile}
\altaffiltext{12}{Institut de Ciències de l'Espai (IEEC-CSIC)}
\altaffiltext{13}{The Carnegie Institute for Science; Carnegie Observatories}
\altaffiltext{14}{University College London}
\altaffiltext{15}{Tel Aviv University}
\altaffiltext{16}{The Ohio State University}
\altaffiltext{17}{INAF, Osservatorio Astronomico di Bologna}
\altaffiltext{18}{JPL, California Institute of Technology}
\altaffiltext{19}{INAF, Osservatorio Astronomico di Trieste}
\altaffiltext{20}{European Southern Observatory}
\altaffiltext{21}{Universitas Sternwarte, M\"unchen}

\begin{abstract}
Growing observational evidence now indicates that nebular line emission has a significant impact on the rest-frame optical fluxes of $z\sim5-7$ galaxies observed with Spitzer. This line emission makes $z\sim5-7$ galaxies appear more massive, with lower specific star formation rates.  However, corrections for this line emission have been very difficult to perform reliably due to huge uncertainties on the overall strength of such emission at $z\gtrsim5.5$.  Here, we present the most direct observational evidence yet for ubiquitous high-equivalent width (EW) [OIII]+H$\beta$ line emission in Lyman-break galaxies at $z\sim7$, while also presenting a strategy for an improved measurement of the sSFR at $z\sim7$.   We accomplish this through the selection of bright galaxies in the narrow redshift window $z\sim6.6-7.0$ where the IRAC 4.5 micron flux provides a clean measurement of the stellar continuum light. Observed 4.5 micron fluxes in this window contrast with the 3.6 micron fluxes which are contaminated by the prominent [OIII]+H$\beta$ lines.  To ensure a high S/N for our IRAC flux measurements, we consider only the brightest ($H_{160}<26$ mag) magnified galaxies we have identified in CLASH and other programs targeting galaxy clusters.  Remarkably, the mean rest-frame optical color for our bright seven-source sample is very blue, $[3.6]-[4.5]=-0.9\pm0.3$.   Such blue colors cannot be explained by the stellar continuum light and require that the rest-frame EW of [OIII]+H$\beta$ be greater than 637{\AA} for the average source.  The bluest four sources from our seven-source sample require an even more extreme EW of 1582{\AA}.   Our derived lower limit for the mean [OIII]+H$\beta$ EW could underestimate the true EW by $\sim2\times$ based on a simple modeling of the redshift distribution of our sources.  We can also set a robust lower limit of $\gtrsim\rm 4\,Gyr^{-1}$ on the specific star formation rates based on the mean SED for our seven-source sample.  Planned follow-up spectroscopy of our sample and deeper IRAC imaging with the SURF'S Up program will further improve these results.

\end{abstract}

\section{Introduction}
\label{sec:intro}
In the last decade the evolution of galaxies at the earliest times has been predominantly mapped out by studying the 
rest-frame UV light in galaxies across cosmic time \citep[e.g.][]{Stanway2003,Bouwens2007,Bouwens2011,Lorenzoni2011,Oesch2012,Oesch2013,Bradley2012b,Bowler2012,Schenker2013b}. Despite great progress in this area, an equally important part of the story regards the build-up of mass in galaxies and the specific star formation rate (sSFR, i.e. the star formation rate divided by the stellar mass), which provide direct constraints on the growth time scale of individual sources \citep{Stark2009,Gonzalez2010}. Typical sSFRs of star-forming galaxies at $z\sim2$ ($M_\ast\sim 5\times10^9M_\odot$) are $\rm\sim2\,Gyr^{-1}$, equivalent to a doubling time of $\sim500\rm\,Myr$.

Over the last few years, there has been a substantial improvement in our characterization of the sSFR in high-redshift
galaxies and how it evolves. 
Initial observational studies found little evolution in the sSFR from $z\sim2$ to higher 
redshift in apparent disagreement with theories of star formation fueled by cold accretion (Stark et al. 2009; Gonzalez et al. 2010; Labbe et al. 2010a,b). However, the effect of nebular emission lines (e.g., [OIII], [OII], H$\alpha$) that can contaminate the 
IRAC measurements of the stellar continuum light had not been taken into account \citep[e.g.][]{Schaerer2009,Schaerer2010}. 

The effect of this emission on broadband IRAC measurements can be quite considerable. Extrapolating the H$\alpha$ EWs measured by \citet{Fumagalli2012} and \citet{Erb2006} to higher redshifts suggests H$\alpha$ EWs as large as 1000{\AA} at $z\gtrsim6$. This would indicate that $\sim$45\% of the flux in [4.5] is due to H$\alpha$  for galaxies at $z\sim6-7$, while [OIII]+H$\beta$ can contribute $\sim$55\% of the flux in [3.6]. 
Correcting for the effects of nebular emission, one can derive sSFRs which are plausibly consistent with theoretical
expectations \citep{Stark2013,Gonzalez2012,deBarros2012}.

As the previous discussion indicates, it is essential in quantifying the sSFR at $z>5$ to characterize the EWs of nebular emission lines and their impact on the IRAC photometry.
Pioneering studies in the last two years have quantified the strength of nebular emission lines at $z\gtrsim4$, through the measured flux offsets to the Spitzer/IRAC [3.6] and [4.5] bands. 
\citet{Shim2011} compare the [3.6] and [4.5] fluxes at $z\sim4$ and show that the $[3.6]-[4.5]$ color correlates with the star formation rate (SFR), implying that the source of the offset is likely due to the presence of H$\alpha$ emission lines. \citet{Stark2013} estimate the influence of H$\alpha$ on the [3.6] flux at $z\sim3-4$ by comparing the color distribution of contaminated and uncontaminated spectroscopic confirmed galaxies \citep[see also][]{Schenker2013b} and extrapolating the observed emission line contamination to $z\sim5-7$. 

The first attempt to derive nebular line EWs for a large sample of Lyman-break galaxies at $z\gtrsim5$ is presented in \citet{Labbe2012}, based on a comparison of a stacked [3.6] and [4.5] flux measurement at $z\sim8$ from the IRAC Ultra Deep Field (IUDF) program with similar flux measurements from a stacked sample at $z\sim7$ (see also \citealt{Gonzalez2012a} who make inferences about the EWs of nebular emission lines from the stacks of $z\sim4-6$ galaxies). Estimates of the nebular-line EWs have also been made from direct fits to large number of spectroscopic ally-confirmed $z\sim4-7$ galaxies \citep{deBarros2012,Ono2012,Tilvi2013,Curtis2013}

Even making use of the above methods, the sSFR in $z\sim6-8$ galaxies is still very
uncertain.   While one can certainly estimate the sSFR in this redshift range by  utilizing an extrapolation of the H$\alpha$ EWs found at $z\sim4$ to higher redshift, extrapolations are inherently uncertain. Results on the sSFR  at $z\sim8$ \citep{Labbe2012}, though providing good leverage to constrain the redshift evolution, are limited by the extreme faintness of the individual galaxies whose redshift distribution is only approximately known.  
Finally, the typical H$\alpha$ EW in $z\sim4$ galaxies used for sSFR estimates has been established primarily through sources which
show Ly$\alpha$ in emission; however, it is unclear if those sources are representative of the broader $z\sim4$ population (for more discussion see \citealt{Schenker2013b}).

To overcome these issues, here we make use of a new strategy for measuring 
the sSFRs and stellar masses for galaxies at very high redshifts, while
simultaneously obtaining very good constraints on the EWs of [OIII]+H$\beta$ line emission.  Our plan is to take
advantage of the considerable quantity of deep, wide-area observations over the 524-orbit, 25-cluster 
Cluster Lensing And Supernova survey with Hubble (CLASH) program \citep{Postman2012} and other programs observing strong lensing clusters with deep multiband HST data. We select a small sample of bright, magnified galaxies for which we can obtain a clean measurement of the stellar continuum light from the deep IRAC observations over these clusters.  
One particularly fruitful redshift window in which we can obtain such clean measurements is the redshift window $z\sim6.6-7.0$, where [4.5] is completely free of any emission lines.  This should allow us to place much more robust constraints on the sSFR and the EW of nebular emission of star-forming galaxies at $z\sim7$.

This paper is organized as follows. In \S\ref{sec:Observations} we discuss our data set, our photometric procedure, and source selection. In \S\ref{sec:Results} we present the properties of our selected $z\sim7$ sample. We discuss the constraints we put on the EWs of H$\alpha$, H$\beta$ and [OIII] and the sSFR. We present a summary and discussion of our results in \S\ref{sec:Summary}.

Throughout this paper we adopt a Salpeter IMF with limits 0.1-100$\,M_\odot$ \citep{Salpeter}. For ease of comparison with previous studies we take $H_0=70\,\rm km\,s^{-1}\,Mpc^{-1},\,\Omega_{\rm{m}}=0.3\,$and$\,\Omega_\Lambda=0.7$. Magnitudes are quoted in the AB system \citep{OkeGun}

\section{Observations}
\label{sec:Observations}
\subsection{Data}
In selecting our small sample of bright, magnified $z\sim7$ galaxies, we make use of the deep HST observations available over the first 23 clusters in the CLASH multi-cycle treasury program (GO \#12101: PI Postman), Abell 1689 and Abell 1703 (GO \#11802: PI Ford), the Bullet cluster (GO \#11099: PI Bradac), and 9 clusters from the Kneib et al. (GO \#11591) program.  The CLASH cluster fields are each covered with 20-orbit HST observations spread over 16 bands using the Advanced Camera for Surveys (ACS: $B_{435},\,g_{475},\,V_{606},\,r_{625},\,i_{775},\,I_{814},$ and $z_{850}$), Wide Field Camera WFC3/UVIS	($UV_{225},\,UV_{275},\,U_{336}$ and $U_{390}$) and WFC3/IR instrument ($Y_{105},\,J_{110},\,J_{125},\, JH_{140}$ and $H_{160}$). Abell 1703 was covered with 22 orbits of ACS and WFC3/IR ($B_{435},\,g_{475},\,V_{606},\,r_{625},\,i_{775},\,z_{850},\,J_{125},\,H_{160}$) while clusters in the Kneib et al. program were covered with 6 orbits ($I_{814},\,J_{110},\,H_{160}$).
HST mosaics were produced using the Mosaicdrizzle pipeline (see \citealt{Koekemoer2011} for further details), and individual bands in the deep imaging data reach 5$\sigma$ depths of 26.4-27.7 mag (0.4"-diameter aperture).  Deep Spitzer/IRAC observations of our fields in the [3.6] and [4.5] bands were provided for by the ICLASH (GO \#80168: \citealt{Bouwens2011b}) and  Spitzer IRAC Lensing Survey program (GO \#60034: PI Egami).   The typical exposure time per cluster was 3.5 to 5 hours per band, allowing us to reach 26.5 mag at 1 sigma.
Reductions of the IRAC observations used in this paper were performed with MOPEX \citep{Makovoz2005}.

\begin{figure}
\centering
\includegraphics[width=0.9\columnwidth,trim=25mm 0mm 70mm 20mm]{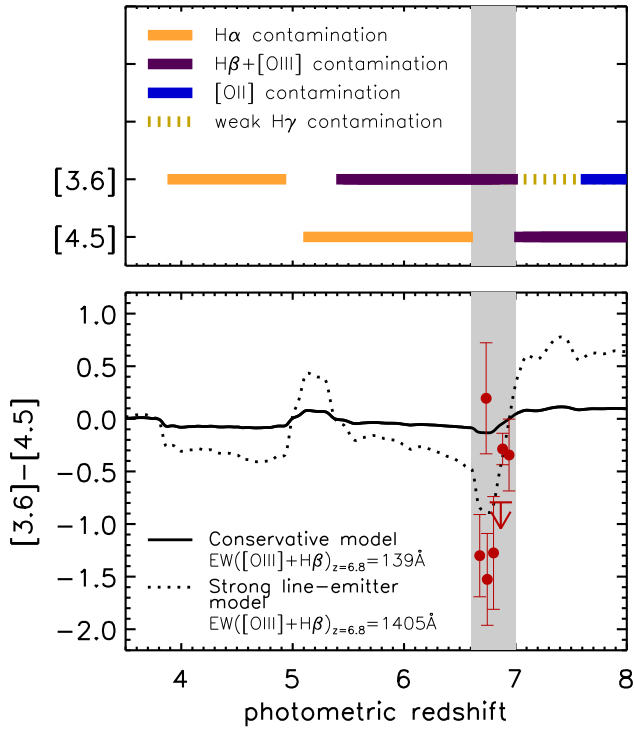}
\caption{The impact of emission lines on the [3.6] and [4.5] band fluxes and our strategy for deriving 
sSFRs and [OIII]+H$\beta$ EWs from our $z\sim7$ sample.   
\textit{Top panel:} The redshift range over which strong nebular emission lines, H$\alpha$, H$\beta$, [OIII] and [OII], will contaminate the [3.6] and [4.5] flux of galaxies.   
\textit{Bottom panel:} The expected $[3.6]-[4.5]$ colors as a function of redshift due to nebular 
emission lines.   The solid and dotted lines show the expected color assuming relatively low EWs, i.e.,
EW$_0$([OIII]+H$\beta)\sim140$\AA, and assuming strong evolution, i.e., EW$_0$([OIII]+H$\beta)\propto(1+z)^{1.8}${\AA} \citep{Fumagalli2012}, respectively, similar to the models considered in \citet{Gonzalez2012} and \citet{Stark2013}.  We select sources in the redshift range 
$z_{phot}=6.6-7.0$, where [OIII]$\lambda\lambda$4959,5007 and H$\beta$ are present in [3.6], while [4.5] receives
no significant contamination from nebular emission lines, falling exactly in between the H$\alpha$ and [OIII] lines. The red solid circles and 1
sigma upper limit show the observed colors in our sample. We find that most sources show blue  $[3.6]-[4.5]$ colors,  
falling in the range between our two models. Four sources from our sample exhibit extremely blue rest-frame optical 
colors, with $[3.6]-[4.5]\lesssim-0.8$, indicating contamination of [OIII]+H$\beta$ with a mean EW of $\gtrsim1582${\AA} (see \S\ref{sec:stack}), even higher than using the \citealt{Fumagalli2012} extrapolation indicated by the dotted line. Two sources at $z\sim6.75$ have been offset by $\Delta z=0.05$ for clarity. }
\label{fig:emlines}
\end{figure}

\begin{figure}
\centering
\includegraphics[width=0.9\columnwidth,trim=10mm -5mm 10mm 0mm]{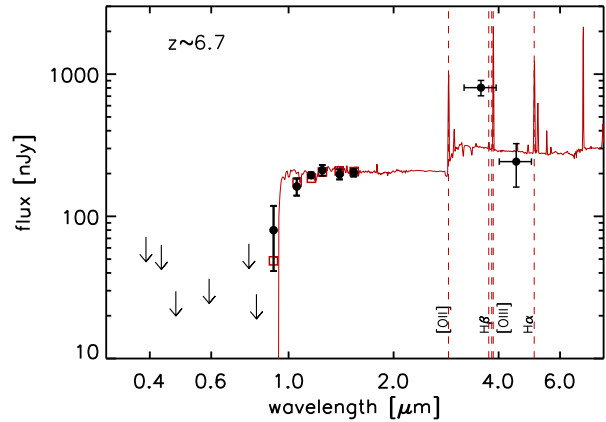}
\caption{The observed HST+Spitzer fluxes (black circles) and model 
spectral energy distribution (red) for one $z\sim7$ candidate rxj1347Z-7362045151 that exhibits a very blue 
$[3.6]-[4.5]$ color.  Because of the brightness of sources in our sample and the many HST filters with deep
observations in the CLASH program, their photometric redshifts are well determined.  This is important 
for establishing that our selected sources are likely in the target redshift window. The [3.6] and [4.5] bands that are shown are not used for the photo-z determination in order to avoid coupling the selection of our sources to the $[3.6]-[4.5]$ colors.  }
\label{fig:emlines_example}
\end{figure} 

\subsection{Photometry and Selection}
\label{sec:phot}

The photometry we obtain for sources in our cluster fields follows a similar procedure as described in \citet{Bouwens2012}.  In short, we run the SExtractor software \citep{Bertin1996} in dual-image mode.  The detection images are constructed from all bands redwards of the Lyman break  (i.e. $Y_{105},\,J_{110},\,J_{125},\, JH_{140}$ and $H_{160}$).  After PSF-matching the observations to the $H_{160}$-band PSF, colors are measured in Kron-like apertures and total magnitudes derived from 0.6"-diameter circular apertures.   

 Our initial source selection is based on the Lyman-break technique \citep{Steidel1999}, with the requirement that the source drops out in the $I_{814}$ band. Specifically, our requirements for $z\sim6-7$ sources are 
\[(I_{814}-J_{110}>0.7)\,\wedge\,(J_{110}-JH_{140}<0.45).\]
For sources in the CLASH program we require $H_{160}<26$ AB, while we select sources to the brighter magnitude limit $H_{160}<25$ AB in all other fields to ensure good photometric redshift constraints for all our sources. We also require sources to have either a non-detection in the $V_{606}$ band ($<2\sigma$) or to have a very strong Lyman break, i.e. $V_{606}-J_{125}>2.5$.  We require sources to be undetected in the optical $\chi^2$ image \citep{Bouwens2011} we construct from the observations bluewards of the $r_{625}$ band. 
 Finally we require the SExtractor stellarity parameter  (equal to 0 and 1 for extended and point sources, respectively) in the $ J_{110}$ band be less than 0.92  to ensure that our selection is free of contamination by stars.  

To identify those sources where we can obtain clean rest-frame optical stellar continuum, we also require that sources have a best-fit photometric redshift between $z=6.6$ and $7.0$, as determined by the photometric redshift software EAZY \citep{Brammer2008}. All available HST photometry (i.e. 16 bands for CLASH clusters) is used in the redshift determinations.  No use of the Spitzer/IRAC photometry is made in the photometric redshift determination to avoid coupling the selection of our sources to the $[3.6]-[4.5]$ colors we will later measure. We use templates of young stellar populations with no Ly$\alpha$ emission. 

Strong Ly$\alpha$ emission can systematically influence the photometric redshift estimate. However, we emphasize that any potential sources from outside our desired redshift interval that could be in our sample due to uncertainties in the photometric redshift estimate would only serve to increase the flux in the [4.5] band and redden the $[3.6]-[4.5]$ color (i.e. due to contamination in the [4.5] band of H$\alpha$ at $z<6.6$ and [OIII]+H$\beta$ at $z>7.0$). Correcting for this possible source of interlopers would result in higher EWs and sSFRs than in the case of no contamination. This reinforces the point we will make in \S\ref{sec:Results} that the EWs we derive for the [OIII]+H$\beta$ emission and the sSFRs are strong lower limits on the actual values.

Figure \ref{fig:emlines}  shows the redshift range where we would expect the strongest emission lines, H$\alpha$, H$\beta$, [OIII]$\lambda\lambda4959,5007$ and [OII]$\lambda3727$, to impact the [3.6] and [4.5] fluxes.
The top panel indicates which lines fall in specific IRAC filters at a given redshift, while the bottom panel indicates the estimated $[3.6]-[4.5]$ color offset due to the various emission lines. We select sources in the redshift range $z_{phot}=6.6-7.0$, where we know that both [OIII] and H$\beta$ fall in [3.6], while [4.5] falls exactly between [OIII] and H$\alpha$ where no significant emission lines are present (see for example Figure \ref{fig:emlines_example}).

\begin{figure}
\centering
\includegraphics[width=1.\columnwidth,trim=0mm 0mm 0mm 0mm]{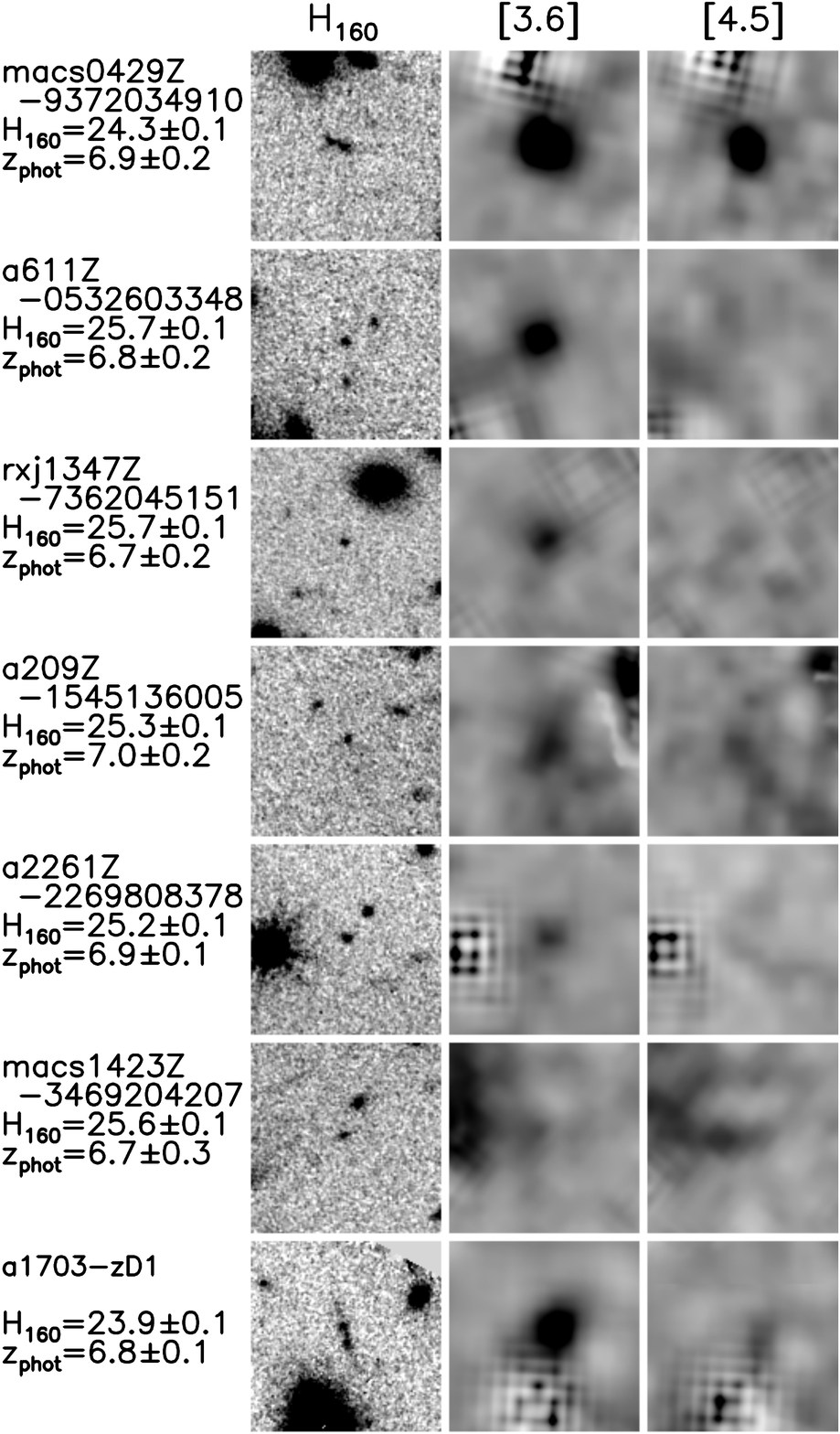}
\caption{HST $H_{160}$, Spitzer/IRAC [3.6], and [4.5] postage stamp images (6.5" $\times$ 6.5") of our sample of 
bright, magnified $z\sim6.6-7.0$ galaxies behind clusters. The IRAC postage stamps have 
already been cleaned for contamination from neighboring sources (\S\ref{sec:iracphot}). It is obvious that a large fraction of the sources in our selection  are much brighter at 3.6$\mu$m than at 4.5$\mu$m.}
\label{fig:exampleclean}
\end{figure}

\begin{figure*}
\centering
\includegraphics[width=0.6\textwidth,trim=0mm 0mm 0mm 0mm]{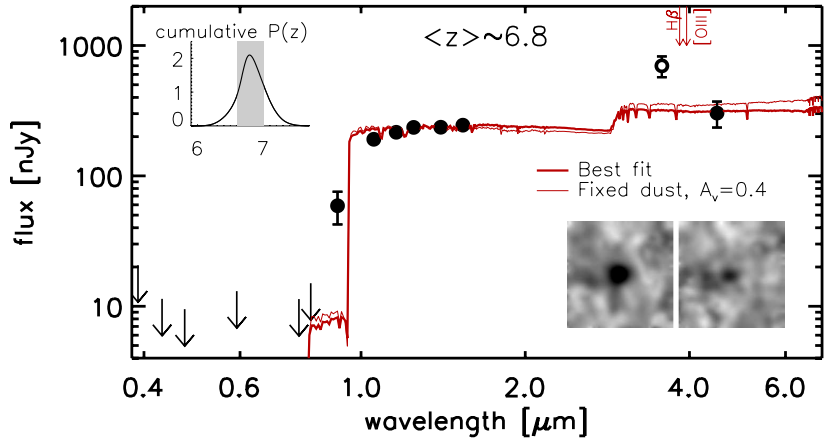}
\caption{The mean-stacked SED of our $z\sim6.6-7.0$ sample.   The mean $[3.6]-[4.5]$ color for our
sample is $\sim-0.9$ and indicates significant line contamination for the typical $z\sim7$ galaxy.  
A small fraction of the sources in our photometric redshift selection are expected to lie outside the target range $z\sim6.6-7.0$
(due to redshift uncertainties).  Any such contamination would make the $[3.6]-[4.5]$ colors redder due to the contribution of H$\alpha$ (at $z<6.6$) or [OIII]+H$\beta$ (at $z>7$) to the [4.5] flux. The thick red line indicates the best fit SED to our observed photometry, excluding the [3.6] flux from the fit.  The thin red line indicates a best fit SED when the dust content is fixed to $A_V=0.38\pm0.16$, similar to the assumptions used in \citet{Bouwens2012}, \citet{Stark2013}, \citet{Gonzalez2012} and \citet{Labbe2012}. }
\vspace{0.2cm}
\label{fig:results_sed}
\end{figure*}

\subsection{IRAC Photometry}
\label{sec:iracphot}

Photometry of sources in the available Spitzer/IRAC data over our fields is challenging, due to blending with nearby sources from the broad PSF.   We therefore use the automated cleaning procedure described in  \citet{Labbe2010a,Labbe2010b}.  In short, we use the high-spatial resolution HST images as a template with which to model the positions and flux profiles of the foreground sources.  The flux profiles of individual sources are convolved to match the IRAC PSF and then simultaneously fit to all sources within a region of $\sim$13" around the source.  Flux from all the foreground galaxies is subtracted and photometry is performed in 2.5"-diameter circular apertures. We apply a factor of $\sim2.0\times$ correction to account for the flux outside of the aperture, based on the radial light profile of the PSF. Figure \ref{fig:exampleclean} shows the cleaned IRAC images of our sample.  Our photometric procedure fails when contaminating sources are either too close or bright.  Sources with badly subtracted neighbors are excluded. In total, clean photometry is obtained for 78\% of the sources, resulting in 7 sources in our final selection (excluding only one source behind RXJ1347 and one source behind MACS1206 from our sample).

\begin{deluxetable*}{lrrccccccc}
\tablewidth{0cm}
\tablecaption{$z\sim6.6-7.0$ candidates included in this work.\label{tab:clusters}}
\tablehead{\colhead{ID} & \colhead{RA} & \colhead{Dec} &  \colhead{$z_{phot}$} & \colhead{$H_{160}$} & \colhead{$H_{160}-[4.5]$} & \colhead{$[3.6]-[4.5]$} & \colhead{$\beta$}  & \colhead{$\mu^{\rm a}$} & \colhead{$M_{\rm UV}^{\rm b}$}}
\startdata
MACS0429Z-9372034910 & $04{:}29{:}37.20$ & $-2{:}53{:}49.10$ & $6.9 \pm 0.2$ & $24.3 \pm 0.1$ & $0.7 \pm 0.1$ & $-0.3 \pm 0.1$ & $-1.4 \pm 0.4$ & $2.5 \pm 0.2$ & $-21.6 \pm 0.1$ \\
A611Z-0532603348 & $08{:}00{:}53.26$ & $36{:}03{:}34.8$ & $6.7 \pm 0.2$ & $25.7 \pm 0.1$ & $0.3 \pm 0.4$ & $-1.5 \pm 0.4$ & $-1.5 \pm 0.5$ & $1.8 \pm 0.1$ & $-20.5 \pm 0.1$ \\
RXJ1347Z-7362045151 & $13{:}47{:}36.20$ & $-11{:}45{:}15.1$ & $6.7 \pm 0.2$ & $25.7 \pm 0.1$ & $0.2 \pm 0.4$ & $-1.3 \pm 0.4$ & $-2.2 \pm 0.5$ & $2.7 \pm 0.2$ & $-20.1 \pm 0.1$ \\
A209Z-1545136005 & $01{:}31{:}54.51$ & $-13{:}36{:}00.5$ & $6.9 \pm 0.2$ & $25.3 \pm 0.1$ & $1.0 \pm 0.3$ & $-0.3 \pm 0.3$ & $-2.7 \pm 0.6$ & $1.2 \pm 0.0$ & $-21.4 \pm 0.1$ \\
A2261Z-2269808378$^{\rm c}$ & $17{:}22{:}26.99$ & $32{:}08{:}37.8$ & $6.9 \pm 0.1$ & $25.2 \pm 0.1$ & $<$0.0 & $<-$0.8  & $-2.0 \pm 0.3$ & $5.6 \pm 1.7$ & $-19.9 \pm 0.3$ \\
MACS1423Z-3469204207 & $14{:}23{:}46.92$ & $24{:}04{:}20.7$ & $6.7 \pm 0.3$ & $25.6 \pm 0.1$ & $1.0 \pm 0.3$ & $0.2 \pm 0.5$ & $-1.2 \pm 0.8$ & $4.7 \pm 1.0$ & $-19.6 \pm 0.3$ \\
A1703-zD1$^{\rm d}$ & $13{:}14{:}59.41 $ & $ 51{:}50{:}00.8 $ & $6.8 \pm 0.1$ & $23.9 \pm 0.1$ & $0.2 \pm 0.2$ & $-1.3 \pm 0.5$ & $-1.4 \pm 0.3$ & $9.0 \pm 4.5$ & $-20.6 \pm 0.5$ \\
\hline
Mean stack & & & $6.8\pm0.2$ & $25.5\pm0.1$ & $0.2\pm0.2$ & $-0.9\pm0.3$ & $-1.9\pm0.3$ & & 
\enddata
\tablenotetext{a}  {The lens models for RXJ1347, MACS0429, Abell 611, Abell 2261 and Abell 209 are made with an improved version of the method described in  \citet{Zitrin2009} and will be published in Zitrin et al. (in prep). The model for Abell 2261 is described \citet{Coe2012}. The model for  MACS1423 is described in \citet{Zitrin2011}, but here we use a refined model (the CLASH collaboration, in prep). The model for Abell 1703 is described in \citet{Zitrin2010}. Errors on the magnification factors are  typical errors at a given $\mu$, calculated based on uncertainty modeling of Abell 383, Abell 611, MS2137 and MACS1423 (Zitrin et al., in prep.).}
\tablenotetext{b}{The intrinsic UV magnitude is derived from the H-band magnitude, corrected for the magnification, $\mu$. The quoted uncertainty includes the estimated uncertainty in $\mu$. }
\tablenotetext{c}{While this source is very compact, our size measurements indicate that it is slightly extended.}
\tablenotetext{d}{Previously reported in \citet{Bradley2012a}}
\end{deluxetable*}

\section{Results}
\label{sec:Results}

Our search for bright ($H_{160}\lesssim26$) LBGs in the redshift range $z\sim6.6-7.0$ behind strong lensing clusters results in 9 candidates. 
One of the sources in our $z\sim7$ sample was previously reported by \citet{Bradley2012a} based on a study of Abell 1703.
For seven sources we obtain reasonably clean IRAC photometry, as shown in the postage stamps in Figure \ref{fig:exampleclean}.  The properties of the sources are summarized in Table \ref{tab:clusters} and they range in $H_{160}$ band magnitude from 24.3 to 25.7. Typical magnification factors, $\mu$, for our sources are $\sim2-9$, using the lensing models of \citet{Zitrin2010,Zitrin2011} and Zitrin et al. (in prep). Though the magnification of the sources improves the S/N of our measurements, we stress that measurements of emission line EWs and sSFRs only depend on the colors of the SED and therefore are not impacted by uncertainties in the model magnification factors.

\begin{figure}
\centering
\includegraphics[width=0.9\columnwidth,trim=20mm 10mm 60mm 30mm]{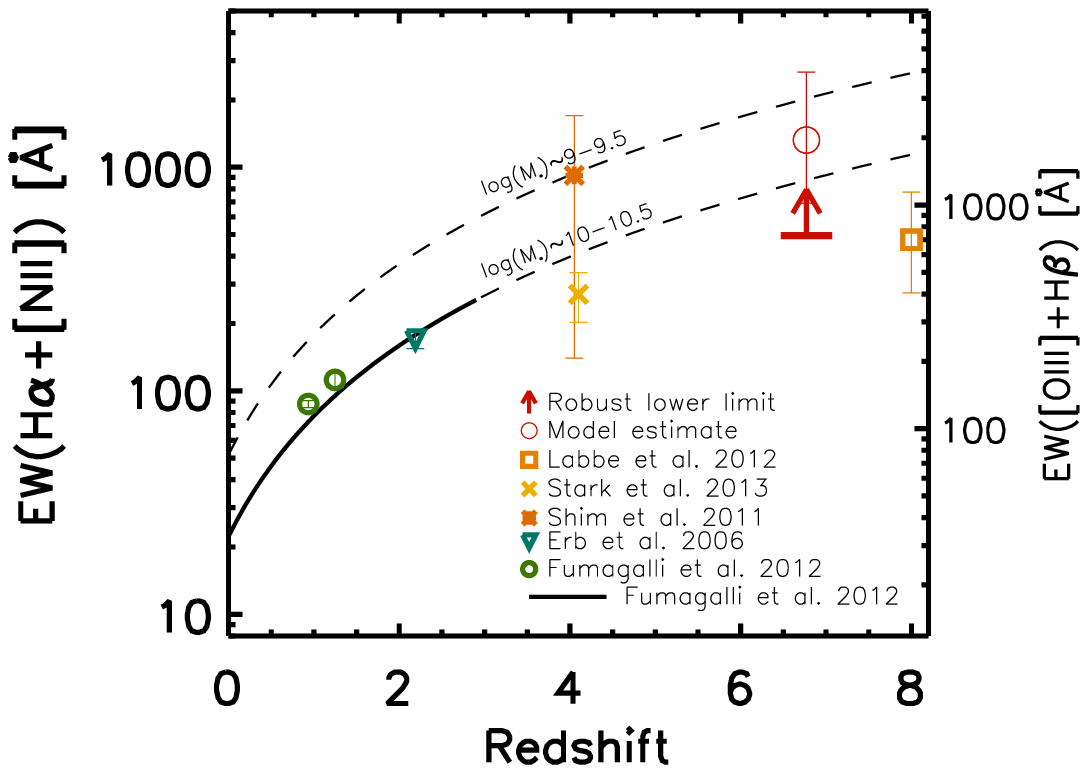}
\caption{ The constraint on the evolution of [OIII]+H$\beta$ EWs (and equivalent  H$\alpha$ EWs) from our stacking analysis and references from the literature \citep{Erb2006,Shim2011,Fumagalli2012,Stark2013,Labbe2012}. The robust lower limit (red arrow) assumes all sources are at $z=6.76$ where the [3.6]-[4.5] color is  expected to be the most extreme for a given set of EWs and that the underlying stellar continuum has a $[3.6]-[4.5]$ color of $\sim-0.4$ (which would only be the case if all galaxies have an age of $\sim3\times10^{6}\rm yr$). Any bluer $[3.6]-[4.5]$ color would therefore arise from the impact of the [OIII]+H$\beta$ emission lines on the [3.6] flux. For the model estimate (red open circle) we model the effects of a broader redshift distribution as described in \S\ref{sec:stack}. For comparison with lower redshift estimates we have converted our EWs to EW(H$\alpha$+[NII]) using the conversion factors from \citet{Anders2003}.
The high EW inferred from our mean stacked  sample indicates significantly stronger emission lines than observed at redshift $z\sim0-2$, possibly consistent with an extrapolation of the trends with redshift and mass found by \citet[indicated by the dashed black lines]{Fumagalli2012}. }
\vspace{0.2cm}
\label{fig:results_EW}
\end{figure}

\subsection{$[3.6]-[4.5]$ color distribution and nebular emission lines}
\label{sec:EmLines}
Our selection of sources in the redshift range $z\sim6.6-7.0$ provides us with the valuable opportunity to establish the typical EW of the nebular emission lines in $z\gtrsim6$ sources through a comparison of the flux in [3.6] and [4.5]. LBGs at high redshift are expected to exhibit flat optical stellar continuum, based on stellar population synthesis models. In these models young galaxies with typical ages between 50-200Myr and low dust extinction, e.g. E(B-V)$\sim0.1$, will have a ([3.6]-4.5])$_{\rm continuum}$ color of $\sim0\pm0.1$ mag. However, extremely young (i.e. $\sim3\times10^6\rm yr$), dust-free galaxies can exhibit ([3.6]-4.5])$_{\rm continuum}$ colors as blue as $\sim-0.4$. To be conservative, we will adopt this for the color of the underlying stellar continuum, and assume that any bluer $[3.6]-[4.5]$ color arises from the impact of emission lines to establish robust lower limits.

 In the bottom panel of Figure \ref{fig:emlines} the dotted line shows a prediction of the observed optical color due to emission lines for a model of strongly increasing rest-frame emission line EWs as a function of redshift (dotted line), with EW$_0$([OIII]+H$\beta)\propto(1+z)^{1.8}${\AA}, based on the evolution in EW$_0$(H$\alpha$) found by \citet{Fumagalli2012} for star forming galaxies over the redshift range $0\lesssim z\lesssim2$. 
The red points show the observed colors for our sample. Most of our sources show quite blue $[3.6]-[4.5]$ colors and essentially all of them are bluer than that expected based on a conservative model of constant rest-frame EW (solid black line: i.e. assuming no evolution from $z\sim2$ where EW$_0$([OIII]+H$\beta$)$\sim$140{\AA}, derived from the H$\alpha$ EWs found by \citealt{Erb2006}). Interestingly enough, three of the sources from our sample have $[3.6]-[4.5]$ colors even bluer than expected at $z\sim6.7-6.8$ for the model from \citet{Fumagalli2012} with EW$_0$(H$\alpha)\propto(1+z)^{1.8}${\AA}. Four of the sources have $[3.6]-[4.5]$ colors bluer than $-0.8$. Since we would only expect galaxies to show such sources extreme $[3.6]-[4.5]$ colors in the narrow redshift range  $z\sim6.6-7.0$, this provides us with additional confidence that our selection is effective at identifying sources in the desired redshift range.

\subsection{Inferred [OIII]+H$\beta$ EWs of $z\sim7$ galaxies from the mean SED}
\label{sec:stack}

To obtain our best measurement of the  [4.5] flux and hence stellar continuum light from $z\sim7$ galaxies,  we use a mean stack of the clean [3.6] and [4.5] images after dividing by the observed rest-frame UV luminosity (the geometric mean of the $J_{125},\,JH_{140}$ and $H_{160}$ luminosities). We measure the flux in a 2.5" diameter aperture on the stacked image and apply an aperture correction measured from the PSF images ($\sim2.0\times$). The mean SED of our stacked $z\sim6.8$ sample is shown in Figure \ref{fig:results_sed}. Errors are obtained through bootstrap resampling. 

We use the stacked detections in the IRAC bands to evaluate the mean contribution of the emission lines. From the mean $[3.6]-[4.5]$ color we estimate the [OIII]+H$\beta$ EW by assuming that our entire sample is at $z=6.76$, where we expect the most extreme colors because [4.5] is completely free of emission lines, while [3.6] is contaminated by both the [OIII] doublet and H$\beta$. In practice, this results in an underestimate of the intrinsic line strength, since we know that the [OIII] lines start to drop out of [3.6] at $z\sim6.9-7.0$. Therefore we expect a less extreme $[3.6]-[4.5]$ color for a given mean EW at $z\sim6.9-7.0$ than at  $z\sim6.7-6.8$. 
It is also possible that due to uncertainties in the photometric redshifts, sources outside of our target redshift range have been included in our selection and therefore the measurement of the [4.5] flux is contaminated by either H$\alpha$ ($z<6.6$) or [OIII] ($z>7$).  This would also make the mean $[3.6]-[4.5]$ color redder and accordingly make the  emission lines appear to be less extreme.

 The mean observed $[3.6]-[4.5]$ color for our sample is $-0.9\pm0.3$ (error obtained through bootstrap resampling). In the most conservative estimate, we assume that the underlying stellar continuum exhibits a $[3.6]-[4.5]$ color of $\sim-0.4$ and therefore the [OIII] and H$\beta$ are responsible for a color of  $[3.6]-[4.5]\sim-0.5$, which would give a robust lower limit of EW$_0$([OIII]+H$\beta)\gtrsim637${\AA} for the mean $z\sim7$ galaxy distribution. The [OIII]+H$\beta$ EW we estimate here is equivalent to EW$_0$(H$\alpha$+[NII])$\gtrsim495${\AA} adopting the tabulated values from \citet{Anders2003} for 0.2Z$_\odot$ metallicity and assuming case B recombination.

 The mean observed $[3.6]-[4.5]$ color for our four bluest sources is $-1.4\pm0.4$.  If we assume again a very blue underlying continuum (i.e., $-0.4$), line emission would be responsible for a color of $[3.6]-[4.5]\sim-1.0$, consistent with a robust lower limit of EW$_0$([OIII]+H$\beta)\gtrsim1582${\AA} for these sources.   While the four bluest sources in our sample are, given their extreme colors, almost certainly at a redshift $z\sim6.7-6.8$, it is unclear if the three other sources are less extreme due to a lower [OIII]+H$\beta$ EW or simply because they lie in a different redshift range (i.e. $z\gtrsim6.9$ or close to $z\sim6.6$), where for a given EW we expect somewhat redder colors.
 
To obtain a good estimate of the EW, detailed knowledge of the redshift distribution and the underlying stellar continuum color is required. Given our lack of deep spectroscopy for our sample we will make some reasonable assumptions to obtain a model estimate of the rest-frame EW of our sample. 
We assume that the redshift probability distribution of our sources is given by the sum of the probability distributions obtained from our photometric redshift code, corrected for the fact that galaxies are more difficult to observe at higher redshift scaling roughly as $\frac{d\log{\phi}}{dz}=0.3$ \citep{Bouwens2012b} and assuming no sources are outside our desired redshift range $z\sim6.6-7.0$. We use this redshift probability distribution in our sample to estimate the expected $[3.6]-[4.5]$ color distribution, for a given mean EW$_0$([OIII]+H$\beta)$ and 0.3 dex scatter around the mean. We assume the underlying continuum color is $-0.25$, as would be expected for a galaxy age of $\sim$10Myr. We randomly draw sources from the distribution and calculate the 68\% likelihood of finding a mean $<[3.6]-[4.5]>$ color given a total of seven observed sources and the observed photometric errors. Based on this modeling, the observed $[3.6]-[4.5]\sim-$0.9 mag color is consistent with a possible EW$_0$([OIII]+H$\beta)$ EW of $\sim 1806_{-863}^{+1826}${\AA}. This is equivalent to EW$_0$(H$\alpha+\rm [NII])\sim 1323^{+1338}_{-632}${\AA}, adopting the same conversion factor from \citet{Anders2003} assumed above.

The modeling we perform above indicates that the true EW$_0$([OIII]+H$\beta)$ may be $\sim 2-3\times$ larger than our robust lower limit of 637{\AA}.
Figure \ref{fig:results_EW} compares our results with other determinations from the literature \citep{Erb2006,Shim2011,Fumagalli2012,Stark2013,Labbe2012}. 
The solid and dashed lines in Figure \ref{fig:results_EW} show the expected evolution of the H$\alpha$+[NII]
EWs extrapolating the evolution found in \citet{Fumagalli2012} at $z\sim0-2$.
 We note however that a direct comparison is difficult to make since the \citet{Fumagalli2012} relation was derived for galaxies in the mass range $M_\ast=10^{10}-10^{10.5} M_\odot$, while we are probing galaxies in the mass range $10^{9}-10^{9.5} M_\odot$. For reference we show the possible evolution of galaxies $M_\ast=10^{9}-10^{9.5} M_\odot$ (top dashed line), using the same scaling with redshift EW$_0$(H$\alpha$+[NII]$)\propto(1+z)^{1.8}${\AA} but extrapolating the normalization to lower masses, based on the mass trend in the SDSS-DR7 data derived in \citet{Fumagalli2012}.

 In general, the EWs we infer are in good agreement with extrapolations from previous results at lower redshift. 
However our results are based on a UV-selected sample, which could yield different results from a mass complete sample. It is clear nonetheless, that our EWs estimates strongly support the high EWs used by \citet{Stark2013} and \citet{Gonzalez2012} in correcting the SEDs of $z\sim5-7$ samples to derive higher values of the sSFRs.


\begin{figure}
\centering
\includegraphics[width=0.9\columnwidth,trim=20mm 10mm 70mm 30mm]{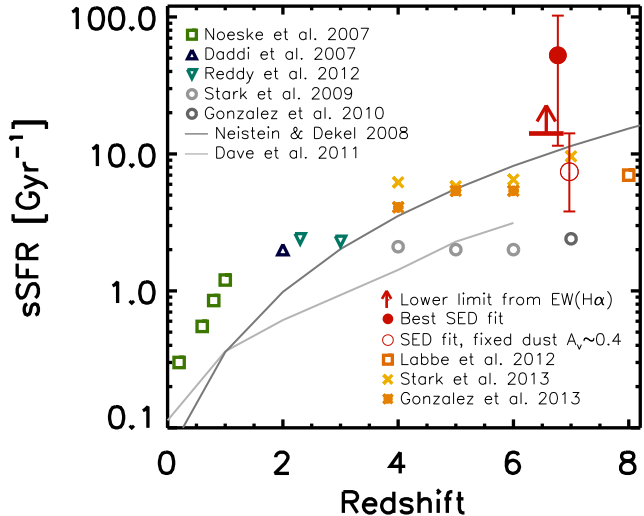}
\caption{  The evolution of the sSFR as a function of redshift, based on the fitted SED to our stacked spectrum, 
excluding the [3.6] flux from the fit. Error bars are the 68\% confidence interval, based on the photometric uncertainties. The red filled circle indicates the best fit using the assumptions described in \S\ref{sec:ssfr}, leaving dust as a free parameter. The red open circle indicates a fit where the dust content is fixed at $A_V=0.38\pm0.16$. We also show a lower limit (red arrow) on the sSFR we derive from the EW of [OIII]+H$\beta$, by converting to H$\alpha$ assuming the line ratios from \citet{Anders2003}. For context, we also include many previous sSFR results from the literature \citep{Noeske2007,Daddi2007,Stark2009,Stark2013,Gonzalez2010,Gonzalez2012,Reddy2012,Labbe2012}.   Our results indicate possible strong evolution in the sSFRs from  redshift $z\sim2$ to $z\sim7$, consistent with other recent results based on an extrapolation of the $z\sim4$ H$\alpha$ EW distribution \citep{Stark2013,Gonzalez2012} or for single $z\sim7$ galaxies \citep[e.g.][]{Ono2012,Tilvi2013,Ouchi2013}.   Our results are also in agreement with theoretical predictions (e.g., \citealt{Neistein2008} [dark grey line] and \citealt{Dave2011} [light grey line]).}
\vspace{0.2cm}
\label{fig:results_ssfr}
\end{figure}

\subsection{Specific Star Formation Rates}
\label{sec:ssfr}
The redshift range where we select galaxies is the only redshift window at $z\gtrsim5$ where we can probe the rest-frame stellar continuum light in an uncontaminated fashion, using the [4.5] IRAC band (see figure \ref{fig:emlines}). This allows us to estimate the sSFR, with minimal contamination from emission lines. 

We obtain the mean sSFR through stellar population modeling of our stacked photometry, leaving out the [3.6] measurement. The modeling was performed with FAST \citep{Kriek2009}, using the \citet[hereafter BC03]{Bruzual2003} stellar populations synthesis models. We use a \citet{Salpeter} IMF with limits 0.1-100$\,M_\odot$ and a \citet{Calzetti2000} dust-law. We consider ages between 10Myr and the age of the universe at $z\sim6.8$ and dust extinction between $A_V=0-2$. 
A constant star formation history and subsolar metallicity (0.2Z$_\odot$) is assumed.
We fix the redshift to the median of the photometric redshifts, at $z=6.77$. 

Given this freedom of parameters the mean SED is best described by a fairly young galaxy (age $\lesssim100$Myr) and reasonable dust ($A_V\sim0.7$) in order to fit both the small Balmer break ($H_{160}-[4.5]\sim0.2$) and moderately red UV-continuum slope ($\beta\sim-1.9$), resulting in a notably high sSFR of $\rm 52^{+50}_{-41}\,Gyr^{-1}$ (see the SED in Figure \ref{fig:results_sed}). However, the interpretation of this result is not straightforward. First of all, we have only the [4.5] band in the rest-frame optical to break the age-dust degeneracy. Given our modest sample there is still a range of models that can fit the data well. A possibly more insightful answer is obtained when we fix the dust to the expected value derived from the typical spread of UV-continuum slopes and the \citet{Meurer1999} law \citep[e.g.][]{Bouwens2012} similar to the assumptions made in \citet{Bouwens2012}, \citet{Stark2013}, \citet{Gonzalez2012} and \citet{Labbe2012}. This results in a dust content of $A_V=0.38\pm0.16$ using the latest numbers from \citet{Bouwens2013}. The fit is shown with the thin red line in Figure \ref{fig:results_sed}. We obtain a sSFR of $\rm 7^{+7}_{-3}\,Gyr^{-1}$. 

Alternatively, we can estimate the sSFR of our $z\sim7$ sample from the [OIII]+H$\beta$ EWs we infer, by converting to H$\alpha$ EW assuming same line ratios from \citet{Anders2003} as described in \S\ref{sec:stack}. We use the \citet{Kennicutt1998} relation to convert H$\alpha$ luminosity to star formation rate and we use BC03 models (assuming no dust) to convert the rest-frame optical continuum light to stellar mass. Using these assumption we obtain a lower limit on the sSFR for our seven source sample and the bluest four sources of $\sim\rm14\,Gyr^{-1}$ and  $\sim\rm130\,Gyr^{-1}$ respectively, based on the robust lower limits on the [OIII]+H$\beta$ EW derived in \S\ref{sec:stack}.



Comparing with direct constraints at $z\sim2$ we estimate $\gtrsim 2\times$ evolution in the sSFR over this redshift range, in good agreement with estimates at $z\sim7$ based on a few spectroscopically confirmed sources \citep{Ono2012,Tilvi2013} and
extrapolating the H$\alpha$ EWs from lower redshifts \citep{Stark2013,Gonzalez2012}. Our derived constraint is also in agreement with theoretical models that predict the sSFR to follow the specific infall rate of baryonic matter \citep[e.g.][]{Neistein2008}.

\section{Summary and Discussion}
\label{sec:Summary}

In this paper, we present the cleanest evidence yet for very high [OIII]+H$\alpha$ EWs in the $z\sim7$ galaxy population.   We also
simultaneously explore a strategy for obtaining a clean measurement of the sSFR at $z\sim7$ based on the stellar continuum
flux measured in the [4.5] micron band -- which is largely free of contamination from the strongest nebular lines.
Nebular emission lines ([OIII], H$\alpha$, H$\beta$) and the extreme faintness of $z\gtrsim5.5$ galaxies make it extremely challenging 
to establish the stellar masses and sSFRs of the typical galaxy at high redshift.

To overcome these issues, we have isolated a small sample of nine bright ($H_{160} < 26$ mag), magnified galaxies in the redshift range $z\sim6.6-7.0$ from CLASH and other programs, seven of which we can perform high-quality IRAC photometry.  Galaxies with photometric redshifts in the range $z\sim6.6-7.0$ are useful, since there the [4.5] band from Spitzer/IRAC provides us with a clean measurement of the stellar continuum flux from galaxies in the rest-frame optical, free of contamination from dominant nebular emission lines (Figure \ref{fig:emlines} and Figure \ref{fig:emlines_example}). 

For the mean source in our sample, we find that we can set a robust lower limit on the rest-frame EW of
[OIII]+H$\beta$ of 637{\AA}.   For this lower limit, we adopt the bluest conceivable $[3.6]-[4.5]$ colors for the 
stellar continuum and assume that all sources in our sample are at $z=6.76$ where a given [OIII]+H$\beta$ EW would 
produce the most extreme $[3.6]-[4.5]$ color.   Use of a more realistic redshift distribution for our sample, i.e.,
consistent with the photometric redshift estimates and not assuming that all sources are at $z=6.76$, suggest
that these lower limits may underestimate the true EWs by a factors of $\sim2\times$.

The four bluest sources in our selection (58\% of our sample) show evidence for even more extreme line 
emission, with $[3.6]-[4.5]\lesssim-0.8$.   For these 4 sources, we can set a robust lower limit of 1582{\AA} on the rest-frame EW in [OIII]+H$\beta$.  

Extreme line emission with EWs greater 1000{\AA} has been found at lower redshift in low mass galaxies \citep{vanderwel2011,Atek2011}. Our results are consistent with the idea that extreme line emission may be present in the typical star-forming galaxy at $z\sim7$. 

Furthermore, our [4.5] stack results imply a firm lower limit on the sSFR of $\sim\rm 4\,Gyr^{-1}$ for star-forming 
galaxies at $z\sim7$.   If any sources from our $z\sim6.6-7.0$ photometric redshift sample lie at lower or
higher redshifts than this, it would imply even lower [4.5] micron fluxes for the stack and hence higher
sSFRs. Compared with sSFRs measurements at $z\sim2$ \citep{Daddi2007,Reddy2012}, this implies at least a $\gtrsim 2\times$ evolution in the sSFR over the redshift range $z\sim2$ to $z\sim7$.   Similar to a few other spectroscopically confirmed $z\sim7$ galaxies in the literature \citep{Ono2012,Tilvi2013}, this provides strong evidence that the sSFRs at $z\sim7$ are high.

We expect improvement in these results through the measurement of spectroscopic redshifts for our sample from deep spectroscopy.  This should allow us to obtain an even cleaner selection of $z\sim6.6-7.0$ galaxies from which to quantify the emission line contamination and sSFRs.
Follow-up observations of our sample are facilitated by the fact that these candidates are typically $\sim$1 magnitude brighter than similar candidates found in the field, making these efforts quite feasible in terms of the telescope time required.

Moreover, our bright $z\sim7$ sample is small and the S/N we have per source is still modest. Increases in sample size can come from shallow surveys over a larger numbers of clusters, such as those available from recent snapshot programs. S/N increases will come from
very deep HST+Spitzer observations being taken by the Frontier Fields program and the SURF'S Up program \citep{Bradac2012}.

\acknowledgments
We thank  Jeff Cooke, Rob Crain, Eichii Egami, Andrea Ferrara, Marijn Franx, Max Pettini, Norbert Pirzkal and Vivienne Wild for interesting conversations. Eichii Egami independently discovered the same extreme $[3.6]-[4.5]$ colors in at least  one of the sources from the present sample. We thank Pascal Oesch for useful feedback on our manuscript. We acknowledge support from ERC grant HIGHZ \#227749, an NWO vrij competitie grant, 
and the NASA grant for the CLASH MCT program. AZ is supported by contract research ``Internationale Spitzenforschung II/2-6'' of the Baden W\"urttemberg Stiftung.

\bibliographystyle{apj}

\end{document}